\documentclass{article}
\usepackage{epsfig}
\usepackage{psfig}
\date{}

\newcommand{\cd}{{\rm cd}}
\newcommand{\dm}{\langle x_0|e^{-\beta\hat{H}}|x_0\rangle}
\newcommand{\cA}{{\cal A}}
\newcommand{\cV}{{\cal V}}
\newcommand{\xgm}{x_{\rm gm}}
\newcommand{\xlm}{x_{\rm lm}}
\newcommand{\xsp}{x_{\rm sp}}
\newcommand{\xct}{x_{\rm ct}}
\newcommand{\xc}{x_{\rm c}}
\newcommand{\qc}{q_{\rm c}}
\newcommand{\qt}{q_{\rm t}}

\begin{document}

\title{\bf Uniform semiclassical approximation
in quantum statistical mechanics\footnote{Work supported
by CNPq, FAPESP, FUJB/UFRJ and the U.S.\ Department of
Energy under Contract No.\ DE-AC02-98CH10886.}}

\author{C.\ A.\ A.\ de Carvalho,$^1$ R.\ M.\ Cavalcanti,$^2$
E.\ S.\ Fraga,$^3$ \\ 
and S.\ E.\ Jor\'as$^4$ \\
\small\it $^1$Instituto de F\'\i sica, 
Universidade Federal do Rio de Janeiro, \\ 
\small\it Caixa Postal 68528, Rio de Janeiro, RJ 21945-970, Brazil \\
\small\it $^2$Instituto de F\'\i sica, Universidade de S\~ao Paulo, \\
\small\it Caixa Postal 66318, S\~ao Paulo, SP 05315-970, Brazil \\
\small\it $^3$Nuclear Theory Group, Physics Department, 
Brookhaven National Laboratory, \\
\small\it Upton, NY 11973-5000, U.S.A. \\
\small\it $^4$High Energy Theory Group, Physics Department, 
Brown University, \\
\small\it Providence, RI 02912, U.S.A.} 

\maketitle

\begin{abstract}

We present a simple method 
to deal with caustics in the semiclassical
approximation to the partition function of a one-dimensional
quantum system. The procedure, which makes use of
complex trajectories, is applied to the quartic double-well
potential.

\end{abstract}


\section{Introduction}

It is well known \cite{Feynman} that
the partition function for a particle of mass $m$ 
interacting with a potential $V(x)$ and a thermal reservoir at 
temperature $T$ can be written as 
a path integral ($\beta=1/k_BT$):
\begin{eqnarray}
\label{Z1}
& &Z(\beta)=\int dx_0\,\dm,
\\
\label{diag}
& &\dm=\int_{x(0)=x_0}^{x(\beta\hbar)=x_0}
[{\cal D}x(\tau)]\, e^{-S[x]/\hbar},
\\
& &S[x]=\int_0^{\beta\hbar} d\tau
\left[\frac{1}{2}\,m\dot{x}^2+V(x)\right].
\end{eqnarray}
The semiclassical approximation to the density
matrix element (\ref{diag}) is given 
by\footnote{Each term in the sum 
on the r.h.s.\ of (\ref{sc}) is in fact the first term
of a series. See Ref.\ \cite{CAAC2}.}
\begin{equation}
\label{sc}
\dm\approx\sum_{k=1}^{N}e^{-S[\xc^k]/\hbar}\,\Delta_k^{-1/2},
\end{equation}
where $\xc^k(\tau)$ is a classical trajectory [i.e.,
it is a solution to the Euler-Lagrange equation,
$m\ddot{x}=V'(x)$, subject to the boundary conditions
$x(0)=x(\beta\hbar)=x_0$] that
{\em minimizes}\footnote{The Euclidean nature of the
path integral allows one to discard saddle-points.}
(globally or locally) 
the action $S[x]$, and $\Delta_k$ is the determinant of 
of the fluctuation operator
$\hat{F}[\xc^k]\equiv -m\,\partial_{\tau}^2+V''[\xc^k]$.
(A derivation of this result will be sketched in 
Section \ref{improved}.)

In Ref.\ \cite{CAAC2} we have examined,
for the sake of simplicity, potentials for which
there is only one classical trajectory satisfying
the above boundary conditions.
In general, however, the number $N$ of such solutions
depends on $x_0$ and $\beta$. A problem then occurs
when we cross a {\em caustic} [the frontier between two regions
of the $(x_0,\beta)$-plane characterized
by different values of $N$]:  
the r.h.s.\ of (\ref{sc}) diverges \cite{CAAC1}.
This divergence, however, is unphysical, being an
artifact of the semiclassical approximation.
The purpose of this work is to present a simple
extension of the semiclassical approximation
which circumvents this problem.\footnote{Ankerhold {\it et al}.\
\cite{Ankerhold} have discussed the caustics problem
near the top of a potential barrier. 
The present work shows how to deal with the problem everywhere.}
(Due to limitations
of space, here we shall only sketch the
method. Details will be given elsewhere.)


\section{Improved semiclassical approximation}
\label{improved}

In order to show how one can improve the semiclassical
approximation so as to eliminate the unphysical
divergences at the caustics, it is convenient to
recall how (\ref{sc}) is derived. Briefly, one has to: 
(i) expand the action around a minimum $\xc(\tau)$:
$S[\xc+\eta]=S[\xc]+S_2+\delta S$, where
$S_2 = \frac{1}{2}\int_0^{\beta\hbar}d\tau\,\eta(\tau)\,
\hat{F}[\xc(\tau)]\,\eta(\tau)$ and
$\delta S=O(\eta^3)$;
(ii) throw away $\delta S$;
(iii) express $\eta(\tau)$ in terms of the orthonormal
modes of $\hat{F}$, i.e.,
$\eta(\tau)=\sum_{j=0}^{\infty}a_j\varphi_j(\tau)$,
where 
$\hat{F}\varphi_j(\tau)=\lambda_j\varphi_j(\tau)$,
$\varphi_j(0)=\varphi_j(\beta\hbar)=0$;
then
$S_2 = \frac{1}{2}\sum_{j=0}^{\infty}\lambda_ja_j^2$
and $ [{\cal D}x(\tau)]
=\prod_{j=0}^{\infty}da_j/\sqrt{2\pi\hbar}$.
The path integral in (\ref{diag}) has now become
a product of Gaussian integrals. Performing the
integrations one arrives at
the ``usual'' semiclassical approximation
to the density matrix element:
\begin{equation}
\label{exp1}
\dm\approx e^{-S[\xc]/\hbar}\,\Delta^{-1/2},
\end{equation}
where $\Delta=\prod_{j=0}^{\infty}\lambda_j={\rm det}\,\hat{F}$.
If there are $N$ minima, one has to add together their 
contributions, thus obtaining (\ref{sc}).

When we cross a caustic, a
classical trajectory $\xc(\tau)$ is created or annihilated.
Precisely at this point, the lowest eigenvalue of
$\hat{F}[\xc]$ vanishes, thus making the integral 
$\int_{-\infty}^{\infty} da_0\,\exp(-\lambda_0a_0^2/2\hbar)$ 
diverge.
This problem can be remedied by retaining fluctuations
beyond quadratic in the subspace 
spanned by $\varphi_0$ (the eigenmode of $\hat{F}$ 
associated with $\lambda_0$). As a result of this 
procedure,\footnote{The procedure adopted here 
resembles the treatment of caustics in optics \cite{Berry}
and in quantum mechanics \cite{DV}, suitably modified
to take into account the Euclidean nature of the path
integral in (\ref{diag}).}
we obtain an improved approximation
to the density matrix element (\ref{diag}):
\begin{equation}
\label{unif}
\dm\approx e^{-S[\xgm]/\hbar}\,\Delta^{-1/2}
\,{\cal F}(x_0,\beta),
\end{equation}
where $\xgm(\tau)$ is the {\em global minimum} of $S[x]$ and
\begin{equation}
\label{calF}
{\cal F}(x_0,\beta)\equiv\sqrt{\frac{\lambda_0}{2\pi\hbar}}
\int_{-\infty}^{\infty} da_0\,e^{-\cV(a_0)/\hbar},
\end{equation}
with
\begin{equation}
\label{U}
\cV(a_0)=\frac{1}{2}\,\lambda_0a_0^2
+\sum_{n=3}^{M}\left(\int_0^{\beta\hbar}d\tau\,
V^{(n)}[\xgm(\tau)]\,\varphi_0^n(\tau)\right)\frac{a_0^n}{n!}\,.
\end{equation}
A couple of remarks are in order here: 
(i) we take for $M$ [Eq.\ (\ref{U})] the smallest 
even integer such that the coefficient
of $a_0^M$ in $\cV(a_0)$ is 
positive for all values of $x_0$ and $\beta$; this suffices
to make the integral in (\ref{calF}) finite 
even when $\lambda_0$ vanishes;
(ii) the factor $\lambda_0^{1/2}$ in ${\cal F}$
cancels the factor $\lambda_0^{-1/2}$ contained in
$\Delta^{-1/2}$; combined with (i), this shows that
the improved approximation to $\dm$, Eq.\ (\ref{unif}), 
is finite at the caustics; (iii) there is a one-to-one
correspondence between the minima of $S[x]$ and the
minima of $\cV(a_0)$; therefore, it is not 
necessary to explicitly add their contributions as
in (\ref{sc}), for they are already contained
in ${\cal F}$.

Although the procedure outlined above teaches us how to
cross the caustics, it is not very convenient: 
in order to obtain the coefficients of $\cV(a_0)$
one has to find $\lambda_0$ and $\varphi_0(\tau)$.
This, in general, is not an easy
task, and makes the whole procedure very cumbersome.
Instead, we shall present an alternative way of
obtaining those coefficients, which is based on
remark (iii) above.

Let us assume that $M=4$ in Eq.\ (\ref{U}); this is the case
for the quartic double-well potential, to be discussed in
the next section. Then the ``effective action'' 
$\cA(a_0)\equiv S[\xgm]+\cV(a_0)$
for the ``critical'' mode $\varphi_0$ is a
fourth degree polynomial in $a_0$.
Let us assume it has three (real) extrema:
a global minimum at $a_0=0$, a local maximum 
at $u>0$, and a local minimum at $v>u$.
This allows us to write such a polynomial 
as\footnote{One can easily check that $\cA'(0)=\cA'(u)=\cA'(v)=0$.} 
$\cA(a_0)=S[\xgm]+\alpha\left[\frac{1}{2}\,uv\,a_0^2
-\frac{1}{3}\,(u+v)\,a_0^3+\frac{1}{4}\,a_0^4\right]$.
In order to relate the parameters in $\cA(a_0)$ to
calculable quantities, we impose that 
$\cA(v)=S[\xlm]$ and $\cA(u)=S[\xsp]$, where $\xlm(\tau)$
and $\xsp(\tau)$ are the local minimum and the lowest
saddle-point of $S[x]$, respectively.
This yields
\begin{equation}
\label{xi}
\frac{S[\xlm]-S[\xgm]}{S[\xsp]-S[\xgm]}
=\frac{\cA(v)-\cA(0)}{\cA(u)-\cA(0)}
=\frac{\xi^3(2-\xi)}{2\xi-1},
\end{equation}
where $\xi\equiv v/u$.
Another combination of parameters which becomes determined is 
$\mu\equiv\alpha u^4$:
\begin{equation}
\label{lambda}
S[\xsp]-S[\xgm]=\cA(u)-\cA(0)=\frac{\mu}{12}\,(2\xi-1).
\end{equation}
In terms of the calculable parameters 
$\xi$ and $\mu$, 
we can rewrite $\cA(a_0)$ as
$\cA(a_0)=S[\xgm]+\cV_3(a_0/u)$, where
$\cV_3(z)=\mu\left[\frac{1}{2}\,\xi z^2
-\frac{1}{3}\,(1+\xi)\,z^3+\frac{1}{4}\,z^4\right]$.
There still remains one parameter to be determined,
namely $u$. Fortunately, ${\cal F}$ does
not depend on it. Indeed,
identifying $\cV_3(a_0/u)$ with $\cV(a_0)$ 
yields $\lambda_0=\mu\xi/u^2$, 
which allows us to rewrite (\ref{calF}) solely
in terms of $\xi$ and $\mu$: 
\begin{equation}
{\cal F}=\sqrt{\frac{\mu\xi}{2\pi\hbar u^2}}
\int_{-\infty}^{\infty}da_0\,e^{-\cV_3(a_0/u)/\hbar}
=\sqrt{\frac{\mu\xi}{2\pi\hbar}}
\int_{-\infty}^{\infty}dz\,e^{-\cV_3(z)/\hbar}.
\end{equation}

The discussion above can be easily adapted to the case
in which $\cA(a_0)$ has only one extremum. In this case, 
$\cA'(a_0)$ has one real ($a_0=0$) and two
complex conjugate roots ($w$ and $w^*$), the latter corresponding
to the {\em complex} trajectories $\xct(\tau)$ and $\xct^*(\tau)$. 
Accordingly, one has
$\cA(a_0)=S[\xgm]+\cV_1(a_0/|w|)$, where
$\cV_1(z)=\chi\left[\frac{1}{2}\,z^2
-\frac{2}{3}\,(\cos\phi)\,z^3+\frac{1}{4}\,z^4\right]$,
with $\chi\equiv\alpha|w|^4$ and $\phi\equiv{\rm arg}(w)$.
Identifying $\cA(w)$ with $S[\xct]$ then yields
\begin{equation}
S[\xct]-S[\xgm]=\cA(w)-\cA(0)=\frac{\alpha w^3}{12}\,(2w^*-w)
=\frac{\chi}{12}\,\left(2e^{2i\phi}-e^{4i\phi}\right),
\end{equation}
from which we can obtain
$\chi$ and $\phi$. Finally, identifying $\cV_1(a_0/|w|)$  
with $\cV(a_0)$ leads to 
$\lambda_0=\chi/|w|^2$, so that
\begin{equation}
{\cal F}=\sqrt{\frac{\chi}{2\pi\hbar|w|^2}}\,
\int_{-\infty}^{\infty}da_0\,e^{-\cV_1(a_0/|w|)/\hbar}
=\sqrt{\frac{\chi}{2\pi\hbar}}\,
\int_{-\infty}^{\infty}dz\,e^{-\cV_1(z)/\hbar}.
\end{equation}


\section{Application: the quartic double-well potential}
\label{application}

Let us consider the quartic double-well potential,  
$V(x)=\frac{\lambda}{4}\,(x^2-a^2)^2$, $\lambda>0$.
In order to simplify notation, we replace $x$ and $\tau$
by $q\equiv x/a$ and $\theta\equiv\omega\tau$,
respectively, where $\omega\equiv(\lambda a^2/m)^{1/2}$.
In the new variables, the equation of motion reads
$\ddot{q}=U'(q)$, where 
$U(q)=\frac{1}{4}\,(q^2-1)^2$. 
Closed trajectories have the form
\begin{equation}
\qc(\theta)=\qt\,\cd(u,k),
\end{equation}
where $\cd$ is one of the Jacobian elliptic 
functions \cite{GR},
$u\equiv\sqrt{1-\qt^2/2}\,(\theta-\Theta/2)$,
$\Theta\equiv\beta\hbar\omega$, and
$k\equiv\sqrt{\qt^2/(2-\qt^2)}$.
The turning point $\qt$ is fixed by the boundary
condition $\qc(0)=q_0$.
The classical action can be written as 
$S[\xc]=(\hbar/g)\,I[\qc]$, where
$g\equiv\hbar\lambda/m^2\omega^3$ and
\begin{equation}
I[\qc]=\Theta\,U(\qt)+2\,{\rm sgn}(\qt-q_0)
\int_{q_0}^{\qt}\sqrt{2\,[U(q)-U(\qt)]}\,dq.
\end{equation}
Finally, the determinant 
of the fluctuation operator is given 
by \cite{CAAC2} 
\begin{equation}
\label{Delta}
\Delta=4\pi g\,{\rm sgn}(q_0-\qt)\,
\frac{\sqrt{2\,[U(q_0)-U(\qt)]}}
{U'(\qt)}\left(\frac{\partial q_0}{\partial \qt}
\right)_{\Theta}.
\end{equation}

Using these ingredients 
one can compute both the usual and the improved
semiclassical approximations to 
$\langle q_0|e^{-\beta\hat{H}}|q_0\rangle$,
Eqs.\ (\ref{sc}) and (\ref{unif}), respectively.
The results are compared in Fig.\ 1.

\begin{figure}
\begin{center}
\psfig{figure=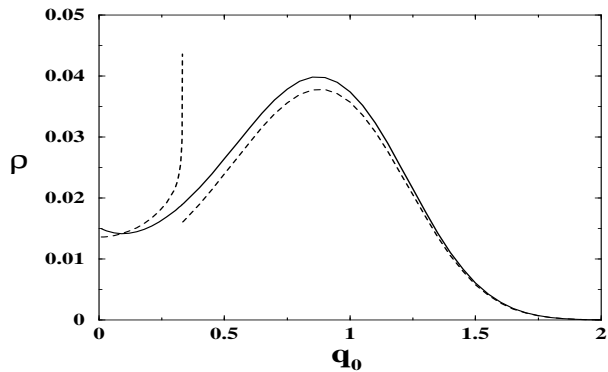,width=8cm,height=5cm}
\caption{$\rho\equiv\langle q_0|e^{-\beta\hat{H}}|q_0\rangle$
{\it vs}.\ $q_0$ for $\Theta=5.0$ and $g=0.3$.
The dashed line is obtained using 
the usual semiclassical approximation [Eq.\ (\ref{sc})]; 
the solid line
is the result of the approximation discussed in 
Section \ref{improved}. The equation $\qc(0)=q_0$ (which
determines the turning point $\qt$) has three real solutions
for $q_0<\tilde{q}_0=0.33319\ldots$; as we cross the caustic,
two of them (the ones associated with a local minimum and
a saddle-point of the action) coalesce --- thus causing the divergence
in the usual semiclassical approximation --- and reemerge at 
the other side as a pair of complex conjugate solutions.}

\end{center}
\end{figure}



\begin{thebibliography}{99}

\bibitem{Feynman} R. P. Feynman, 
{\it Statistical Mechanics} (Addison-Wesley, New York, 1972);
H. Kleinert, 
{\it Path Integrals in Quantum Mechanics, Statistics and 
Polymer Physics} (World Scientific, Singapore, 1995).

\bibitem{CAAC2} C. A. A. de Carvalho, R. M. Cavalcanti, 
E. S. Fraga, and S. E. Jor\'as, 
Ann. Phys. (N.Y.) {\bf 273}, 146 (1999);
Phys. Rev. E {\bf 61}, 6392 (2000). 

\bibitem{CAAC1} C. A. A. de Carvalho and R. M. Cavalcanti,
Braz. J. Phys. {\bf 27}, 373 (1997).

\bibitem{Ankerhold} J. Ankerhold and H. Grabert, 
Physica A {\bf 188}, 568 (1992);
F. J. Weiper, J. Ankerhold, and H. Grabert, 
Physica A {\bf 223}, 193 (1996).

\bibitem{Berry} M. Berry, 
in {\it Physics of Defects}, Les Houches Session XXXV (1980), 
edited by R. Balian, M. Kl\'eman, and J.-P. Poirier 
(North-Holland, Amsterdam, 1981); 
M. V. Berry and C. Upstill, 
in {\it Progress in Optics XVIII}, 
edited by E. Wolf (North-Holland, Amsterdam, 1980).

\bibitem{DV} G. Dangelmayr and W. Veit, 
Ann. Phys. (N.Y.) {\bf 118}, 108 (1979);
L. S. Schulman, 
{\it Techniques and Applications of Path Integration} 
(John Wiley, New York, 1981).

\bibitem{GR} I. S. Gradshteyn and I. M. Ryzhik, 
{\it Table of Integrals, Series and Products} 
(Academic Press, New York, 1965). 

\end{thebibliography}
\end{document}